\documentclass[twocolumn,showpacs,preprintnumbers,amsmath,amssymb]{revtex4}
\usepackage{epsfig}
\usepackage{graphicx}
\usepackage{dcolumn}
\usepackage{bm}
\usepackage{bm}
\def\dis{\displaystyle}
\def\itmb{\begin{itemize}}
\def\itme{\end{itemize}}
\def\enmb{\begin{enumerate}}
\def\enme{\end{enumerate}}
\def\eqnb{\begin{equation}}
\def\eqne{\end{equation}}
\def\PTP{Prog. Theor. Phys.(Kyoto)}

\def\NPB{{Nucl. Phys.} {\bf B}}
\def\PLB{{Phys. Lett.} B}

\def\PRD{{Phys. Rev.} D}

\begin{document}
\newcommand{\ttbs}{\char'134}
\newcommand{\Slash}[1]{\ooalign{\hfil/\hfil\crcr$#1$}}

\title{{\Large Correlation of the ghost and the quark\\
in the lattice Landau gauge QCD}}

\author{Sadataka Furui$^\dagger$ and Hideo Nakajima$^*$}

\affiliation{School of Science and Engineering, Teikyo University, 320-8551 Utsunomiya Japan.$^\dagger$}

\affiliation{Department of Information Science, Utsunomiya University, 321-8585 Utsunomiya Japan.$^*$ }


\begin{abstract}
Effects of the quark field on the ghost propagator of the lattice Landau gauge are investigated by using the unquenched SU(3) configurations produced by the MILC collaboration and compared with quenched gauge configurations of SU(2) first copy and the parallel tempering (PT) gauge fixed samples and quenched SU(3) $56^4$ configurations. 
 We measure the color symmetric and the color antisymmetric ghost propagator
 and the Binder cumulant of the $l^1$ norm and the $l^2$ norm of color antisymmetric ghost propagators and investigate deviation from those of Gaussian distributions. 

In the first copy samples of quenched SU(2) we observe a large fluctuation in the Binder cumulant at the lowest momentum point. This fluctuation is reduced in the PT gauge fixed samples. The color anti-symmetric ghost propagator of quenched SU(3) configurations depends on the lattice size and is small as compared to the symmetric one in the large lattice of $56^4$.

The Binder cumulants of the quenched SU(2) and the unquenched SU(3) are almost consistent with 3-d and 8-d Gaussian distribution, respectively. 
In most cases they are independent of the momentum, but in the case of MILC$_f$ configuration of $\beta_{imp}=7.11$, the Binder cumulant has an anomalous momentum dependence.  This anomaly is correlated with an anomaly of mass function $M(q)$ of the corresponding quark wave function of bare mass $m_0=27.2$MeV.  

A comparison of the SU(3) unquenched configurations and quenched configurations indicates that the dynamical quarks have the effect of making color antisymmetric ghost propagator closer to the Gaussian distribution and the Kugo-Ojima color confinement parameter $c$ closer to 1.

PACS numbers:12.38.Gc, 12.38.Aw, 11.10.Gh, 11.15.Ha, 11.15.Tk, 11.30.Rd

Keyword: Ghost propagator, Quark propagator, Binder Cumulant, Dynamical mass
\end{abstract}

\maketitle

\setcounter{page}{0}

\section{Introduction}
In this presentation, we would like to show our study of color confinement and dynamical chiral symmetry breaking via unquenched lattice Landau gauge simulation using configuration given by the MILC collaboration in which Asqtad Kogut-Susskind (KS) fermions are used.

In our study of color confinement mechanism, we measure 1) the Kugo-Ojima confiment parameter $c$, 2) the $A^2$ condensate in running coupling, gluon propagator and quark propagator and 3) the ghost condensate parameter $v$ and the Binder cumulant of the color anti-symmetric ghost propagator.

We are interested in the effect of quark fields on these variables since the confinement and the chiral symmetry breaking could be related with each other. We show our result of measurement of mass function $M(q)$ of MILC$_c$($20^3\times 64$ lattice) and MILC$_f$($28^3\times 96$ lattice)\cite{MILC}.

In the Kugo-Ojima theory\cite{KO}, unphysical longitudinal gluons and ghosts do not appear in the physical spectrum by the BRST quartet mechanisme.
Using arrows as the BRST transformations and $B$ as the Nakanishi-Lautrup auxiliary field, the BRST transformation in the quark sector can be expressed as
\[
\begin{array}{ccccccc}
\psi & \to & -\psi c & \to & 0 & & \\
      &     & \psi \bar c & \to & -\psi c \bar c-\psi B & \to & 0.
\end{array}
\]
Inclusion of fermion gives more restiction on the degrees of freedom of the ghost and it may change the fluctuation of the ghost propagator.

Landau gauge adopted in this theory suffers from the Gribov problem\cite{Gr}, i.e. the uniqueness of the gauge field is not guaranteed in the simple gauge fixing procedure and in general there are gauge copies. In the Lagrangian field theory, presence of gauge copies is ignored, but in the lattice simulation of the infrared region, importance of the selection of a unique gauge is qualified\cite{Zw, NF,FN04}. The complete  algorithm is however not known, despite various proposals (smearing \cite{HdeF}, parallel tempering (PT)\cite{NF,FN04} etc).

This paper is organized as follows. In sect.II, we review theories of the color confinement in infrared QCD. Results of Kugo-Ojima confinement parameter in lattice  simulation are shown in sect.III. Simulation results of the ghost propagator, the gluon propagator, the QCD running coupling and the quark propagator are shown in sect. IV, V, VI and VII, respectively. Summary and discussion are given in sect. VIII.

\section{Color confinement}

Let us summarize developments of the study of color confinement which are related to our work. 
In 1979 Kugo and Ojima\cite{KO} proposed a color confinement criterion, which will be explained in the following. Gribov\cite{Gr} and Zwanziger\cite{Zw}  gave  condition of the confinement on the infrared exponents of the  gluon propagator and the ghost propagator. 
These exponents define that of the running coupling $\alpha_s(q)$ in $\widetilde{MOM}$ scheme. A measurement of $\alpha_s(q)$ via triple gluon vertex in quenched SU(3) lattice simulation by the Orsay group suggested infrared suppression and a presence of mass-dimension 2, $A^2$ condensates\cite{Orsay05}.

The relation between the mass-dimension 2 condensates and the Zwanziger's horizon condition generated by the restriction of the gauge field in the fundamental modular region was pointed out by the Gent group\cite{Gent}.

Although $A^2$ is not BRST invariant, a mixed condensates with $\bar c c$ is BRST invariant\cite{Kondo}.
The Gent group suggested in the Local Composite Operator (LCO) approach that the $\bar cc$ condensates would manifest itself in the color antisymmetric ghost propagator.

The lattice simulation of the color antisymmetric ghost propagator in SU(2) lattice Landau gauge was performed by the group of Cucchieri\cite{CMM05}.  

\subsection{Kugo-Ojima confinement criterion}

The Kugo-Ojima confinemnt criterion is expressed by the parameter $c$ defined in the two-point function
\begin{eqnarray}
&&(\delta_{\mu\nu}-{q_\mu q_\nu\over q^2})u^{ab}(q^2)\nonumber\\
&&={1\over V}
\sum_{x,y} e^{-iq(x-y)}\langle  {\rm tr}\left({\Lambda^a}^{\dag}
D_\mu \displaystyle{1\over -\partial D}[A_\nu,\Lambda^b] \right)_{xy}\rangle.\nonumber
\end{eqnarray}
 as $u^{ab}(0)=-\delta^{ab}c$ becomes 1. We adopt the SU(3) color matrix $\Lambda$ normalized as $tr\Lambda^a\Lambda^b=\delta^{ab}$.

The parameter $c$ is related to the renormalization factor of the gluon sector,  the ghost sector and the quark sector as
\[
1-c=\frac{Z_1}{Z_3}=\frac{\tilde Z_1}{\tilde Z_3}=\frac{Z_1^\psi}{Z_2}
\]
If the finiteness of $\tilde Z_1$ is proved, divergence of $\tilde Z_3$ is a
sufficient condition. If $Z_3$ vanishes in the infrared, $Z_1$ should have higher order 0. If $Z_2$ is finite $Z_1^\psi$ should vanish.

\subsection{Zwanziger's horizon condition}

Zwanziger proposed realization of the unique gauge by restricting the configuration in the fundamental modular region. The condition on his two-point function given  below is equivalent to the Kugo-Ojima theory in a naive continuum limit. 
\[
\sum_{x,y} e^{-iq(x-y)} \left \langle {\rm tr}\left({\Lambda^a}^{\dag}
D_\mu \displaystyle{1\over -\partial D}(-D_\nu)\Lambda^b\right)_{xy}\right
\rangle
\]
\[
=G_{\mu\nu}(q)\delta^{ab}
=\left(\displaystyle{e\over d}\right)\displaystyle{q_\mu q_\nu\over q^2}\delta^{ab}
-\left(\delta_{\mu\nu}-\displaystyle{q_\mu q_\nu\over q^2}
\right)u^{ab},
\]
Here $e=\left\langle\sum_{x,\mu}{\rm tr}(\Lambda^{a\dag} 
S(U_{x,\mu})\Lambda^a)\right\rangle/\{(N_c^2-1)V\}$,
where $S(U_{x,\mu})$ is the coefficient of $\partial_\mu$ of the covariant 
derivative.
The horizon condition reads $\displaystyle \lim_{q\to 0}G_{\mu\mu}(q)-e=0$,
and the l.h.s. of the condition is 
$
\left(\displaystyle{e\over d}\right)+(d-1)c-e=(d-1)h
$
where $h=c-\dis{e\over d}$ and dimension $d=4$, and it follows that 
$h=0 \to {\rm horizon\  condition}$, and thus the horizon condition coincides
with Kugo-Ojima criterion provided the covariant derivative approaches
the naive continuum limit, i.e., $e/d=1$.\\

\section{Lattice data of the Kugo-Ojima parameter}

In our lattice simulation, we adopt two types of the gauge field definitions:
1) $\log U$ type in which the link variable $U$ and the gauge field are related as $U_{x,\mu}=e^{A_{x,\mu}}$ and 2) $U-$linear type in which $A_{x,\mu}$ is a traceless part of the difference of $U_{x\mu}$ and $U_{x,\mu}^\dagger$ multiplied by one half.

The optimizing function corresponding to the two definitions are
1) The $\l^2$ norm of $A^g$ and 2) $tr(2-({U^g}_{x,\mu}+{{U^g}_{x,\mu}}^\dagger))$, respectively.

Under infinitesimal gauge transformation
$g^{-1}\delta g=\epsilon$, its variation reads for either defintion as 
\[
\Delta F_U(g)=-2\langle \partial A^g|\epsilon\rangle+
\langle \epsilon|-\partial { D(U^g)}|\epsilon\rangle+\cdots,
\]
where the covariant derivativative $D_{\mu}(U)$ for two options reads 
commonly as
\[
D_{\mu}(U_{x,\mu})\phi=S(U_{x,\mu})\partial_\mu \phi+[A_{x,\mu},\bar \phi]
\]
where 
$
\partial_\mu \phi=\phi(x+\mu)-\phi(x)$, and 
$\bar \phi=\dis{\phi(x+\mu)+\phi(x)\over 2}
$
The function $S(U_{x,\mu})$ in the $\log U$ definition is  \\
$S(U_{x\mu})=\frac{adj A_{x,\mu}/2}{{\rm tanh} (adj A_{x,\mu}/2)}$.

 Stationality of $F_U(g)$ means Landau gauge, the local minimum means the Gribov region and the global minimum means the fundamental modular region\cite{Zw}.

In the quenched Landau gauge QCD simulation, the Kugo-Ojima parameter saturated at about 0.8, as shown in TABLE \ref{KO_quench}. The parameter $e/d$ is closer to 1 in the $\log U$ definition of the gauge field.

\begin{table}[htb]
\caption{Kugo-Ojima parameter and Zwanziger parameter in $U-$linear(left) and $\log U$(right). $\beta=6.0$ and $6.4$.}\label{KO_quench}
\begin{center}
\begin{tabular}{c|c|ccc|ccc}
 $\beta$&$L$ &$c_1$ & $e_1/d$ & $h_1$ & $c_2$ & $e_2/d$ & $h_2$ \\
\hline
6.0 &16&  0.576(79) &   0.860(1) & -0.28 & 0.628(94)& 0.943(1) & -0.32\\
6.0 &24&  0.695(63)  &  0.861(1) & -0.17 & 0.774(76)& 0.944(1) & -0.17\\
6.0 &32&  0.706(39)  &  0.862(1) & -0.15 & 0.777(46)& 0.944(1) & -0.16\\
\hline
6.4 &32& 0.650(39) & 0.883(1) & -0.23 & 0.700(42)& 0.953(1) & -0.25\\
6.4 &48& 0.739(65) & 0.884(1) & -0.15(7) & 0.793(61)& 0.954(1) & -0.16\\
6.4 &56& 0.758(52) & 0.884(1) & -0.13(5) & 0.827(27)& 0.954(1) & -0.12\\
\hline
6.45 &56&  &  &  & 0.814(89)& 0.954(1) & -0.14\\
\end{tabular}
\end{center}
\end{table}

We measured the corresponding values of the unquenched configurations of MILC$_c$ and MILC$_f$\cite{FN05b}. The lattice specifications are shown in TABLE \ref{milc_m0}.
\begin{table}[htb]
\begin{center}
\caption {$\beta_{imp}$, the bare quark mass $m_0$, the inverse lattice spacing $1/a$, lattice size and lattice length(fm). Suffices $c$ and $f$ of MILC correspond to coarse lattice and fine lattice. $\beta_{imp}=5/3\times \beta$.}\label{milc_m0}
\begin{tabular}{c|c c c c c c c}
   &$\beta_{imp}$ &  $am_0(ud)$& $am_0(s)$& $1/a$(GeV)&$L_s$ & $L_t$ &$a L_s$(fm)\\
\hline
MILC$_c$ &6.83&       0.040&0.050 & 1.64 & 20 &64&2.41\\
       &6.76&       0.007&0.050 & 1.64 & 20 &64&2.41\\
\hline
MILC$_f$ &7.11&       0.0124&0.031& 2.19 &28 & 96&2.52\\
       &7.09 &       0.0062&0.031 & 2.19 &28 & 96&2.52\\
\hline
\end{tabular}
\end{center}
\end{table}

\begin{table}[htb]
\caption{The Kugo-Ojima parameter for the polarization along the spacial directions $c_x$ and that along the time direction $c_t$ and the average $c$, trace divided by the dimension $e/d$, horizon function deviation $h$ of the unquenched  KS fermion (MILC$_c$,MILC$_f$).  The  $\log U$ definition of the gauge field is adopted. }\label{milc_KO}
\begin{center}
\begin{tabular}{c|c|c|c|c|c|c}
& $\beta_{imp}$ & $c_x$     & $c_t$    &$c$ &  $e/d$        &    $h$     \\
\hline
MILC$_c$&6.76 & 1.04(11)  & 0.74(3) &0.97(16) & 0.9325(1) & 0.03(16)  \\
&6.83 & 0.99(14)  & 0.75(3) &0.93(16) & 0.9339(1) &  -0.00(16) \\
\hline
MILC$_f$&7.09 & 1.06(13)  & 0.76(3) &0.99(17) & 0.9409(1) & 0.04(17)  \\
  &7.11 &1.05(13)   &  0.76(3)& 0.98(17) & 0.9412(1) &  0.04(17) \\
\hline
\end{tabular}
\end{center}
\end{table}

As shown in TABLE \ref{milc_KO}, the Kugo-Ojima parameter of MILC configurations are consistent with 1. The qualitative difference from quenched simulations would be due to the difference of the ghost propagator caused by the quark field.

\section{The ghost propagator}
The ghost propagator is defined as the fourier transform of the matrix element of the inverse Faddeev-Popov operator.

\begin{eqnarray}
FT[D_G^{ab}(x,y)]&=&FT\langle tr ( \Lambda^a \{({\cal  M}[U])^{-1}\}_{xy}
\Lambda^b ) \rangle,\nonumber\\
&=&\delta^{ab}D_G(q^2),  \nonumber
\end{eqnarray}
where ${\mathcal M}=-\partial_\mu D_\mu.$  
The ghost dressing function $G(q^2)$ is defined as $q^2 D_G(q^2)$.

We calculate the overlap to get the color diagonal ghost propagator
\begin{eqnarray}
&&D_G(q)=\frac{1}{N_c^2-1}\frac{1}{V}\nonumber\\
&&\times\delta^{ab}(\langle \Lambda^a\cos{\bf q}\cdot{\bf x}|f_c^b({\bf x})\rangle+\langle \Lambda^a\sin{\bf q}\cdot{\bf x}|f_s^b({\bf x})\rangle)\nonumber
\end{eqnarray}
and color anti-symmetric ghost propagator
\begin{eqnarray}
&&\phi^c(q)=\frac{1}{\mathcal N}\frac{1}{V}\nonumber\\
&&\times f^{abc}(\langle \Lambda^a\cos{\bf q}\cdot{\bf x}|f_s^b({\bf x})\rangle-\langle \Lambda^a\sin{\bf q}\cdot{\bf x}|f_c^b({\bf x})\rangle)\nonumber
\end{eqnarray}
where ${\mathcal N}=2$ for SU(2) and 6 for SU(3). Here ${f_c}^b({\bf x})$ and ${f_s}^b({\bf x})$ are the solution of ${\mathcal M} f^b({\bf x})=\rho^b({\bf x})$ with $\displaystyle \rho^b({\bf x})=\frac{1}{\sqrt V}\Lambda^b\cos{\bf q\cdot x}$ and $\displaystyle \frac{1}{\sqrt V}\Lambda^b\sin{\bf q\cdot x}$, respectively.

\begin{figure}[htb]
\begin{center}
\includegraphics[width=7cm,angle=0,clip]{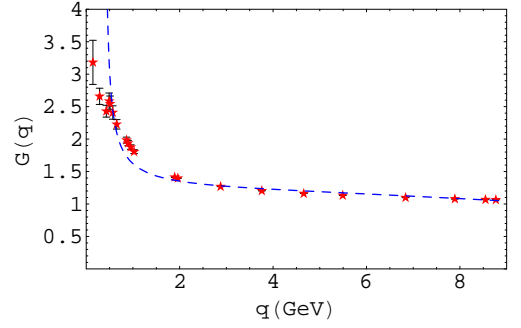}
\end{center}
\caption{The ghost dressing function of MILC$_f$ 
$\beta_{imp}=7.09$(stars) Dashed line is the 4-loop $N_f=2$ pQCD result ($\lambda_G=3.01, y=0.0246100$)}\label{gh_pqcd}
\end{figure}

The ghost dressing function of MILC$_f$ $\beta_{imp}=7.09$ and that of the quenched $\beta=6.45$ $56^4$, the color symmetric and anti-symmetric ghost propagator of MILC$_c$ are shown in \cite{FN06}.  
FIG.\ref{gh_pqcd} is a fit of the ghost dressing function of MILC$_f$ by the perturbative QCD(pQCD) in $\widetilde{MOM}$ scheme\cite{FN05b}.

\subsection{The ghost condensate}
In the LCO approach, a preference of a specific direction of $\langle f^{abc}\bar c^b c^c\rangle$ in the color space is regarded as a signal of the ghost condensate. We define a parameter $v\propto |\langle f^{abc}\bar c^b c^c\rangle|$ and  following \cite{CMM05} measure its value as follows.
We take into account the finite size effect of the lattice through a parameter $r$, defined as
\begin{equation}
\frac{1}{N_c^2-1}\sum_a\frac{L^2}{\cos(\pi \bar q/L)}|\phi^a(q)|=\frac{r}{q^z},\nonumber
\end{equation}
where $L$ is the lattice size and $\bar q=0,1,\cdots L$. In our asymmetric ($L_x< L_t$) lattice and the momentum $q$ near diagonal in the 4-d space, we consider the sine momentum
\begin{equation}
{\tilde q}^2=\sum_{i=1}^3 (2\sin \frac{\pi \bar q_i}{L_x})^2+(2\sin \frac{\pi \bar q_4}{L_t})^2\nonumber
\end{equation}
and choose $L^2$ in the numerator as the square root of the volume $\sqrt{L_x^3 L_t}$ 

The absolute value of the color antisymmetric ghost propagator is parametrized as
\begin{equation}
 \frac{1}{N_c^2-1}\sum_a|\phi^a(q)|= \frac{r/L^2+v}{q^4+v^2}.\nonumber
\end{equation}

\begin{table*}[htb]
\caption{The fitted parameters $r,z$ and $v$ of the color antisymmetric ghost propagator $|\phi(q)|$ of MILC$_c$  and MILC$_f$. The parameters $\bar r,\bar z$ and $\bar v$ are those of the color antisymmetric ghost propagator squared $\phi(q)^2$. Two values of $U$ of MILC$_f$ correspond to the average below $q=1$GeV and the average above 1GeV, respectively.}\label{ghstcdst}
\begin{center}
\begin{tabular}{c|c|c c c | c c c | c}
$\beta_{imp}$ & $m_0$(MeV)& $r$   &$z$ &$v$ & $\bar r$&$\bar  z$& $\bar v$ & $U$\\
\hline
6.76 & 11.5/82.2  &  37.5 & 3.90& 0.012 & 33.5 & 7.6 & 0.045& 0.53(5)\\
6.83 & 65.7/82.2  &  38.7 & 3.85& 0.007 & 33.5 & 7.6 & 0.048& 0.57(4)\\
\hline
7.09 & 13.6/68.0  & 134 & 3.83 & 0.026 & 251 &7.35 & 0.044 &0.57(4)/0.56(1)\\
7.11 & 27.2/68.0  & 112 & 3.81 & 0.028 & 164 &7.34 & 0.002 &0.58(2)/0.52(1)\\
\hline
\end{tabular}
\end{center}
\end{table*}

\begin{figure}
\begin{center}
\includegraphics[width=7cm,angle=0,clip]{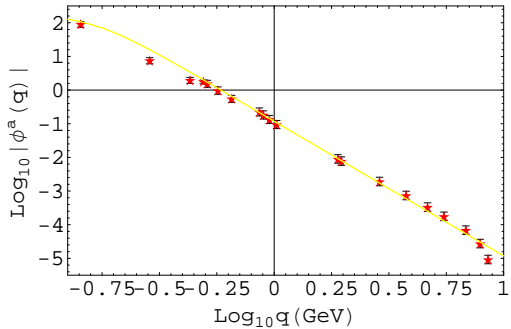}
\end{center}
\caption{$\log_{10}|\vec\phi(q)|$ as the function of  $\log_{10}(q$(GeV)) of MILC$_f$ and the fit using $r=134$ and $v=0.026$GeV$^2$. }\label{phi_709}
\end{figure}

We tried to fit $|\phi(q)|$ and $\phi(q)^2$ of MILC$_f$ $\beta_{imp}=7.09$ and 7.11 and found a relatively good parameter fit of $|\phi(q)|$ with use of $r=134$ and $v=0.026$GeV$^2$, as shown in FIG.\ref{phi_709}. 

Extraction of $v$ from lattice data is not so easy due to the finite size effect of the asymmetric lattice. Except MILC$_f$ $\beta_{imp}=7.09$ which had difficulty in the fitted value from $\vec \phi(q)^2$ is larger than that  from $|\phi(q)|$.

\subsection{Binder cumulant}
Cucchieri et al\cite{CMM05} measured the Binder cumulant of the color antisymmetric ghost propagator
\[
U(q)=1-\frac{\langle\vec\phi(q)^4\rangle}{3\langle\vec\phi(q)^2\rangle^2}.
\]
and found that it is about 0.44, which is compatible with the expectation value of the d-dimensional Gaussian distribution, 
$\displaystyle \frac{\langle \vec\phi^4\rangle}{(\langle \vec\phi^2\rangle)^2}=\frac{d+2}{d}$ with $d=3$.

The Binder cumulant $U(q)$ of the color anti-symmetric ghost propagator $\phi^2(q)$ of MILC$_f$ $\beta=7.09$ $m_0=13.6$MeV/68MeV and MILC$_c$ $m_0=11.5$MeV/82MeV are stable but the data of $\beta_{imp}=7.11$ is rather noisy. In the region $q\geq 1$GeV, the Binder cumulant $U(q)$ is 0.52(1) but in the region $q\leq 1$ GeV it is 0.58(3), which is consistent with that of Gaussian distribution $U(q)\sim 0.583$. The anomalous momentum dependence occurs also in the mass function of the quark propagator, which is discussed in the sect. VII.

We compared the Binder cumulant of SU(2) color antisymmetric ghost propagator of samples which are gauge fixed by the PT method \cite{NF} and the first copy.
The first copy shows large fluctuation at the lowest momentum point. The Binder cumulant of quenched SU(3) seems to depend on the size of the lattice. In the $56^4$ lattice simulation, color anti-symmetric ghost propagator is very weak and the Binder cumulant is noisy.

In \cite{FN05a} we showed the Binder cumulant of MILC$_c$ (21 samples). We increased the statistics of MILC$_c$ to $\beta_{imp}=6.76$ and 6.83, 20 samples each and extended the measurement to MILC$_f$ $\beta_{imp}=7.09$ and 7.11.  In MILC$_c$ $\beta=6.76$, we found an exceptional sample which we encountered also in the quenched SU(3) $56^4$ lattice\cite{FN04}. The ghost propagator of this sample has large exponent $\alpha_G$ which is defined at about $q\sim 0.4$GeV. 
In the case of quenched simulation such copy had larger $\|A_{\mu}\|^2$ than the average and the PT gauge fixing excludes such samples.

\section{The gluon propagator}
The gluon propagator is defined as the two-point function
\begin{eqnarray}
D_{A,\mu\nu}^{ab}(q)&=&\sum_{x={\bf x},t}e^{-iqx}\langle {A_\mu}^a(x){A_\nu}^b(0) \rangle \nonumber\\
&=&(\delta_{\mu\nu}-{q_\mu q_\nu\over q^2})D_A(q^2)\delta^{ab}.\nonumber
\end{eqnarray}
The gluon dressing function is defined as $Z(q^2)=q^2 D_A(q^2)$ 

The gluon propagator and the dressing function of MILC$_f$ $\beta_{imp}=7.09$ are calculated in \cite{FN05b}. 
The data suggest that the gluon propagator is infrared finite. The infrared exponent $\alpha_G$ and it is close to $-2\alpha_G$.  The exponent $\kappa$ of the Dyson-Schwinger equation(DSE) approach is defined at the lower momentum point, inaccessible in the present lattice simulation. 
 We observe the numerical value of $\alpha_G$ is about $\kappa/2$, and the tendency that $\alpha_D$ decreases but $\alpha_D > -1$ i.e. the gluon propagator is not infrared vanishing. The combination of $\alpha_D+2\alpha_G$ is supposed to approach 0 in the DSE approach, but on the lattice it is about -0.1 in the MILC$_c$ and -0.2 in the MILC$_f$.


\section{The QCD running coupling}
In the DSE approach, the ghost-gluon coupling in the $\widetilde{MOM}$ scheme is calculated by the gluon dressing function $Z_3$ and the ghost dressing function $\tilde Z_3$ and the vertex renormalization factor $\tilde Z_1$ as
\begin{equation}
g(q)=\tilde Z_1^{-1} Z_3^{1/2}(\mu^2,q^2) {\tilde Z}_3(\mu^2,q^2)g(\mu)\label{alphagluon}\nonumber
\end{equation}

\begin{figure}[htb]
\begin{center}
\includegraphics[width=7cm,angle=0,clip]{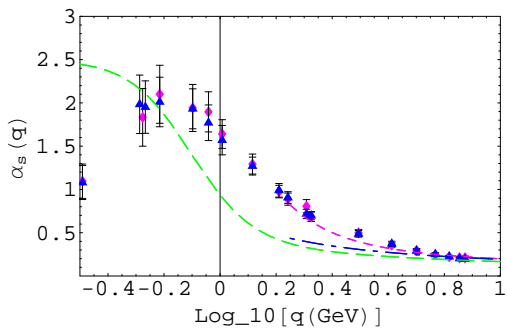}
\end{center}
\caption{The running coupling $\alpha_s(q)$ as a function of  $\log_{10}q$(GeV) of MILC$_c$ ($a=0.12$fm) $\beta_{imp}=6.76$(triangles) and 6.83(diamonds), (50 samles each).}\label{alpmilc}
\end{figure}

Measuring the running coupling from the ghost-gluon coupling is more stable than that of the triple gluon vertex. Our lattice simulation\cite{FN05b} of the gluon propagator and the ghost propagator of MILC$_c$ yield the running coupling shown in FIG.\ref{alpmilc}. There are deviation from the pQCD (dash-dotted line) and the DSE approach with $\kappa=0.5$ (long dashed line).  
As was done by the Orsay group\cite{Orsay05}, we consider a correction including the $A^2$ condensates and obtained $\langle A^2\rangle\sim$ a few GeV$^2$.


The running coupling in the infrared can be estimated from the quark-gluon coupling.
\begin{equation}
g(q)={Z_1^{\psi}}^{-1} Z_3^{1/2}(\mu^2,q^2) Z_2(\mu^2,q^2)g(\mu).\label{alphaquark}\nonumber
\end{equation}
where $Z_2$ is the quark dressing function and ${Z^\psi}_1$ is the vertex renormalization factor. An evaluation of $Z_2(\mu^2,q^2)$ is given in the next section.

\section{The quark propagator}

We extended the measurement of the quark propagator using Asqtad action of MILC$_c$ \cite{FN05a} to MILC$_f$. In the case of MILC$_c$, we compared the Asqtad action and the Staple+Naik action.

Due to long computation time for the convergence of the conjugate gradient method, the number of samples is of the order of 10 for each $\beta_{imp}$ and the bare quark mass $m_0$.

The quark propagator is defined as a statistical average over Landau gauge fixed samples
\[
S_{\alpha\beta}(p)=\left\langle \langle \chi_{p,\alpha}| 
{1\over i\Slash{D}(U)+m_0 }|\chi_{p,\beta}\rangle \right\rangle 
\]
In this expression, the inversion, ${1\over i\Slash{D}(U)+m_0 }$, is performed via conjugate gradient method after preconditioning, and we obtain
\[
S_{\alpha\beta}(q)=Z_2(q)\frac{-i\gamma q+M(q)}{q^2+M(q)^2}
\]
The mass function $M(q)$ reflects dynamical chiral symmetry breaking. In high momentum region, it is parametrized as 
\begin{eqnarray}
M(q)&=&-\frac{4\pi^2 d_M\langle \bar \psi \psi\rangle_\mu [\log (q^2/\Lambda_{QCD}^2)]^{d_M-1}}{3q^2 [\log (\mu^2/\Lambda_{QCD}^2)]^{d_M}}\nonumber\\
&+&\frac{m(\mu^2)[\log (\mu^2/\Lambda_{QCD}^2)]^{d_M}}{[\log (q^2/\Lambda_{QCD}^2)]^{d_M}},\nonumber
\end{eqnarray}
where $d_M=12/(33-2N_f)$  and $m(\mu^2)$ is the running mass. 

In the infrared region, we adopt the monopole fit
\[
M(q)=\frac{\tilde c\Lambda^3}{q^2+\Lambda^2}+m_0
\] 

The momentum dependence of $M(q)$ and $Z_2(q)$ of $m_0=13.6$MeV in the infrared region of Asqtad action is smoother than that of the Staple+Naik action. It could be attributed to the effect of the tadpole renormalization.
The parameters $\tilde c$ and $\Lambda$ in our fit of the mass function are given in TABLE \ref{massfit}.

\begin{table*}[htb]
\caption{The parameters $\tilde c$ and $\Lambda$ of the Staple+Naik action(left) and
the Asqtad action. }\label{massfit}
\begin{center}
\begin{tabular}{c|c|c c c|c c c}
$\beta_{imp}$ & $m_0$(MeV)     & $\tilde c$   &$\Lambda$(GeV) & $\tilde c\Lambda$(GeV) & $\tilde c$   &$\Lambda$(GeV) & $\tilde c\Lambda$(GeV)\\
\hline
6.76 & 11.5  & 0.44(1) & 0.87(2)  & 0.38 &0.45(1) & 0.91(2)& 0.41 \\
     & 82.2  & 0.30(1)  & 1.45(2) & 0.43 & 0.33(1)& 1.36(1) & 0.46\\
6.83 & 65.7  & 0.33(1)  & 1.28(2) & 0.42 & 0.35(1)& 1.25(1) & 0.44\\
     & 82.2  & 0.30(1)  & 1.45(2) & 0.43 & 0.33(1)& 1.34(1) & 0.45 \\
\hline
7.09 & 13.6  & 0.45(1) & 0.82(2) & 0.37  &0.50(2) &0.79(2) &0.39 \\
     & 68.0  &  0.30(1)  & 1.27(4) & 0.38 & 0.35(1)& 1.19(1) & 0.41\\
7.11 & 27.2  &  0.43(1) & 0.89(2) & 0.38 &0.20(2) & 1.04(3) & 0.21\\
     & 68.0  &  0.32(1)  & 1.23(2) & 0.40 & 0.36(1) & 1.15(1) & 0.42\\
\hline
\end{tabular}
\end{center}
\end{table*}

We showed the quark wave function renormalization $Z_\psi(q^2)=g_1(\mu^2)/{\mathcal Z}_2(q^2)$ of MILC$_f$ $\beta_{imp}=7.11$ using the staple+Naik action in \cite{FN05a}, where ${\mathcal Z}_2$ is the bare lattice data and $g_1(q^2)$ is the coefficient of $\gamma_\mu$ of the vector current vertex that compensates the artefact in ${\mathcal Z}_2$. 

 We adopt $\langle A^2\rangle$ as a fitting parameter and calculate\cite{Orsay05}
\begin{eqnarray}
&&Z_\psi(q^2)=\frac{g_1(\mu^2)}{{\mathcal Z}_2(q^2)}\nonumber\\
&&=Z_\psi^{pert}(q^2)
+\frac{\left(\frac{\alpha(\mu)}{\alpha(q)}\right)^{(-\gamma_0+\gamma_{A^2})/\beta_0}}{q^2} \frac{\langle A^2\rangle_\mu}{4(N_c^2-1)}{Z_\psi^{pert}(\mu^2)}\nonumber\\
&&+\frac{c_2}{q^4} \label{Orsayfit}\nonumber
\end{eqnarray}
where $\alpha(q)$ are data calculated in the $\widetilde{MOM}$ scheme using the same MILC$_f$ gauge configuration\cite{FN04}. 

Here $N_f$ is chosen to be 2 but the data does not change much for 3. We choose $\Lambda_{QCD}=0.691$GeV and $\langle \bar \psi\psi\rangle_\mu=-(0.7\Lambda_{QCD})^3$\cite{lat05a,lat05b}.

 Since $g_1(q^2)$ in the infrared is expected to be given by the running coupling, no suppression of the quark wavefunction renormalization suggests that the infrared suppression of the running coupling obtained by the ghost-gluon coupling could be an artefact.

\begin{figure}[htb]
\begin{center}
\includegraphics[width=7cm,angle=0,clip]{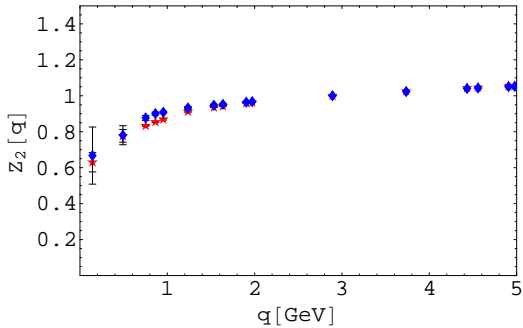}
\end{center}
\caption{The wave function renormalization factor $Z_2(q)$ of MILC$_f$($\beta_{imp}=7.11$ blue diamonds) and MILC$_f$($\beta_{imp}=7.09$ red stars). 
 }\label{z2_709_711}
\end{figure}

In \cite{pbhlwz} the $Z_2(q)$ is normalized to 1 at $q=3$GeV. In our simulation without this kind of renormalization, $Z_2(q)$ at $q=3$GeV is close to 1 and the results are consistent. Our mass function $M(q)$ of $\beta_{imp}=7.09$ are about 20\% larger than those of \cite{pbhlwz}, but if $M(q)$ is renormalized to the theoretical value of \cite{pbhlwz} at $q=3$GeV we reproduce their data. 
Our mass function is consistent with DSE analysis\cite{adfm04}.

In the case of $\beta=7.11$ $m_0=27.2$MeV, the mass function of $\beta_{imp}=7.11$ $m_0=68$MeV is consistent with \cite{pbhlwz}, but that of $m_0=27.2$MeV is suppressed in the infrared and differ from \cite{pbhlwz}. The momentum dependence is correlated with the momentum dependence of the Binder cumulant of color anti-symmetric ghost propagator. The disagreements with \cite{pbhlwz} may be due to Gribov copies. 

\begin{figure}[htb]
\begin{center}
\includegraphics[width=7cm,angle=0,clip]{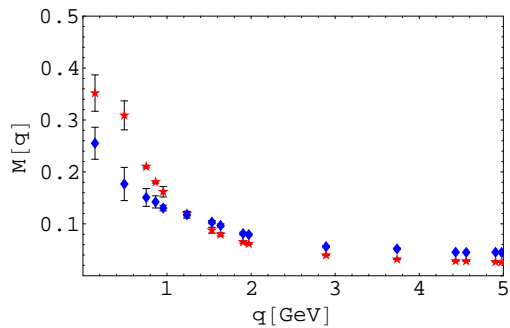}
\end{center}
\caption{The mass function $M(q)$ of MILC$_f$($\beta_{imp}=7.11$ $m_0=27.2$MeV, blue diamonds) and MILC$_f$($\beta_{imp}=7.09$ $m_0=13.6$MeV, red stars). 
 }\label{mass709_711}
\end{figure}

In the case of $\beta_{imp}=7.09$, the chiral symmetry breaking mass measured by using the Staple+Naik action and measured by using the Asqtad action (FIG.\ref{mass709_711}) are consistent within about 5\%, and there is no  anomaly. Whether the Asqtad action has a non-QCD like behavior in a certain parameter region is under investigation.

\begin{figure}[htb]
\begin{center}
\includegraphics[width=7cm,angle=0,clip]{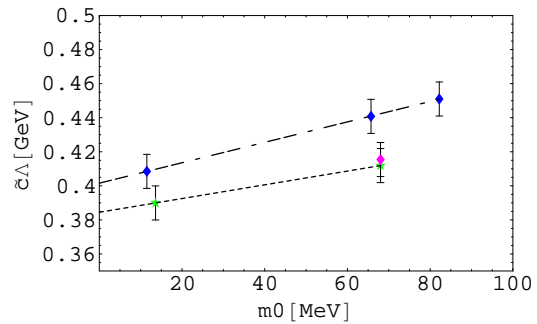}
\end{center}
\caption{The chiral symmetry breaking mass $\tilde c\Lambda$ as a function of bare mass and its chiral limit. Dotted line is  the extrapolation of MILC$_f$ $\beta_{imp}=7.09$ and the dash-dotted line is that of MILC$_c$, Asqtad action. }
\label{chiralmass2}
\end{figure}

\section{Summary and discussion}

In this presentation, we showed our results of unquenched lattice Landau gauge QCD simulation using the KS fermion and infrared features of the ghost propagator, the gluon propagator and the quark propagator. We studied the Kugo-Ojima confinement criterion and the Zwanziger's horizon condition.

The infrared exponent of the ghost propagator $\alpha_G$ of the lattice is about half of the value $\kappa$ of the DSE\cite{SHA, Zw1}. The gluon propagator is infrared finite. The Kugo-Ojima parameter $c$ was saturated at about 0.8 in the quenched SU(3) but became consistent with 1 in the unquenched SU(3).

In $\beta=6.4, 56^4$ quenched SU(3) configurations, we found a copy whose $\alpha_G=0.272$ v.s. $\alpha_G$(average)=0.223, and whose gluon propagator
 has an axis along which the reflection positivity is manifestly violated. 
 In the $\beta_{imp}=6.76, 20^3\times 64$ MILC$_c$ configurations, we find also an exceptional sample.
 The exceptional samples of unquenched configuration causes fluctuation in the Binder cumulant of color antisymmetric ghost propagator in the infrared, similar to the SU(2) first copy. 


In the measurement of the running coupling $\alpha_s(q^2)$ by the ghost-gluon coupling, we observed infrared suppression. The quark wavefunction renormalization $Z_\psi(q)$ in the infrared suggests that the infrared suppression is a lattice artefact and the running coupling in the continuum limit freezes to a finite value. In the analysis, the artefact inherent to the compactness of the manifold should also be taken into account\cite{Fischer}.

In the ghost sector, the Kugo-Ojima confinenment criterion $\displaystyle \frac{\tilde Z_1}{\tilde Z_3}=\frac{1}{\infty}=0$ is satisfied. In the gluon sector, the continuum limit of $Z_3$ in the infrared is unknown, but the finite lattice data suggests that it is finite and non-zero. The infrared limit of $Z_1$ is not known, but the behavior of the ghost propagator suggests that it could be indefinite(fluctuating).

A natural question, 'How do the fixed points of the QCD renormalization group flow behave in the infrared?' may arise.  In 1971, Wilson argued that the fixed point of the renormalization group flow of QCD could have limit cycle structure rather than a fixed point\cite{Wil71}. The deviation of the ghost dressing function from 4-loop pQCD result in the infrared region shown in FIG.\ref{gh_pqcd} suggests that the flow of the ghost wavefunction renormalization is not a trivial one and it is correlated with the presence of dynamical quarks. The theorem on no renormalization effect on the ghost propagator\cite{Tay71} does not guarantee the validity of the simple extension of the ghost wave function renormalization from ultraviolet to infrared.

An extension of the analysis to finite temperature is straightforward. A preliminary simulation using MILC finite temperature configurations\cite{MILC_ft} shows that the Binder cumulant at finite temperature becomes smaller than that of Gaussian distribution as temperature rises. The effect of quenching randomness\cite{LaBi} played by quarks in the zero temperature unquenched configurations becomes weak above the critical temperature $T_c$.  To clarify the nature of the infrared fixed points, systematic studies of finite size effects and the Gribov copy effects\cite{FN04,imss06}  are necessary.

We observed that differences of the color antisymmetric ghost propagator in the quenched/unquenched configuratiions and zero/finite temparature make differences in Kugo-Ojima parameter. The results imply that the quark field and the ghost field are correlated as the BRST quartet mechanism suggests.
\vskip 0.2 true cm
This work is supported by the High Energy Accelerator Research Organization(KEK) supercomputering project No. 05-128 using Hitachi-SR8000 and the large scale simulation program No.5(FY2006) using SR11000.  Numerical calculation of the ghost propagator was carried out by NEC-SX5 at the CMC of Osaka university and by NEC-SX8 at the Yukawa Institute Computer Facility. H.N. is supported by the MEXT grant in aid of scientific research in priority area No.13135210.
S.F. thanks Attilio Cucchieri, Teresa Mendes and Silvio Sorella for the hospitality in Rio.

\end{document}